\documentclass[11pt]{article}
\usepackage{moriond,epsfig}

\usepackage{amsmath,amssymb,amsfonts,amsthm,bbm,cancel,wasysym}
\usepackage{epsfig,graphics,graphicx,epstopdf}
\usepackage{rotating}
\usepackage{colordvi,color}
\usepackage{array,booktabs,colortbl,colordvi,multirow}

\def\Journal#1#2#3#4{{#1} {\bf #2}, #3 (#4)}

\def\be{\begin{equation}}
\def\ee{\end{equation}}
\def\bea{\begin{eqnarray}}
\def\eea{\end{eqnarray}}

\newcommand{\ctoprule}{\toprule[0.5mm]}
\newcommand{\cbottomrule}{\bottomrule[0.5mm]}
\newcommand{\cmrule}{\midrule[0.25mm]}

\newcommand{\crowcolor}{\rowcolor[rgb]{0.9,0.9,0.9}}

\newcommand{\refeq}[1]{\mbox{Eq.~(\ref{#1})}}

\newcommand{\hc}{\mathrm{h.c.}} 

\newcommand{\U}[1]{U(#1)}
\newcommand{\SU}[1]{SU(#1)}

\newcommand{\Adj}{\mathrm{Adj}}

\newcommand{\SM}{\mathrm{SM}}

\begin{document}
\vspace*{4cm}
\title{Z$^\prime$ Bosons and Friends \footnote{Talk at {\em Rencontres de Moriond, EW Interactions and Unified Theories}, Mar 13th-20th, 2011, La Thuile, Italy.}}

\author{ Manuel P\'erez-Victoria}

\address{Departamento de F\'{\i}sica Te\'orica y del Cosmos and CAFPE,\\
Universidad de Granada, E-18071 Granada, Spain}

\maketitle\abstracts{The invariance of extensions of the Standard Model under the full  $\SU{3}_C\otimes \SU{2}_L \otimes \U{1}_Y$ gauge group can be used to classify general vector bosons and to write their interactions in a model-independent fashion. This description is useful for both direct and indirect searches. We comment on electroweak precision limits and show some simple applications to Higgs and top physics.
}

\section{General extra vector bosons and gauge invariance}
New vector bosons are a common occurrence in theories beyond the Standard Model (SM). They appear whenever the gauge group of the SM is extended, as the gauge bosons of the extra (broken) symmetries. This is the case of Grand Unified Theories (GUT), including string constructions, or Little Higgs models. They also occur in theories in extra dimensions, when the gauge bosons propagate through the bulk. 
Strongly-interacting theories, such as technicolor, often give rise to spin 1 resonances. This can be related to the previous possibilities via hidden gauge symmetry or holography. Extra vector bosons are also receiving a lot of attention these days because they are among the best candidates for an early discovery at the LHC.

It is possible to classify vector bosons according to their electric charge: neutral vector bosons, called Z$^\prime$, charge $\pm 1$ vector bosons, called W$^\prime$ and vectors with other integer or fractional charges. On the other hand, the complete theory including the new vectors must be invariant under the full $\SU{3}_C \otimes \SU{2}_L \otimes \U{1}_Y$ gauge group. This imposes additional restrictions on the allowed couplings to the SM fields, and also implies that certain vectors must appear simultaneously and have similar masses. Of course, electroweak symmetry breaking can give rise, in some cases, to splittings in the masses of the different components of a given multiplet. These splittings are of the order of the Higgs vacuum expectation value. 

In Ref.~\cite{Jorge}, Francisco del Aguila, Jorge de Blas and the author have made use of this information to classify the new vectors into irreducible representations of the full SM gauge group, and to study the most general gauge-invariant Lagrangian of this class of SM extensions. Here, we will briefly describe this formalism and show some applications.

The extra vector bosons that have been most extensively studied are neutral singlets, usually associated to an extra abelian gauge symmetry (see, for instance, the review in Ref.~\cite{Langacker:2008yv}). We will go far beyond this particular case, and consider all the representations that could be potentially observable by their indirect effects on precision data or their direct effects at colliders. Our main assumption is that single production of the new vector bosons is possible, so that they have good chances or being observed at large colliders. This requires interactions that couple SM operators to the extra vector fields and are linear in the latter. Since all the leading contributions to electroweak precision data (EWPD) arise from tree-level exchanges of just one heavy vector boson, they are included in this analysis. The only other assumption we make is that the interactions of these extra fields should be renormalizable by power counting, to avoid extra suppressions. From the point of view of a low-energy effective theory, these couplings produce dimension-six operators, while interactions with more than one new vector field in the same operator---and nonrenormalizable interactions---give rise to operators of higher scaling dimension.

The requirement of linear renormalizable couplings, together with Lorentz symmetry and invariance under the full SM gauge group, constrain the possible quantum numbers of the new vectors. In Table~\ref{table:newvectors}, we give the quantum numbers for the 15 irreducible representations of vector fields that can have linear and renormalizable interactions. This table also introduces the notation for each class of vector boson, which is partly inspired by the usual notation for SM fields.
Note that the representations with nonvanishing hypercharge are complex.
\begin{table}[t]
\begin{center}
\caption{Vector bosons contributing to the dimension-six effective Lagrangian. The quantum numbers $(R_c,R_L)_Y$ denote the representation $R_c$ under $SU(3)_c$, the representation $R_L$ under $SU(2)_L$ and the hypercharge $Y$. \vspace{.2cm}}
{\small
\begin{tabular}{ l  c  c  c  c  c  c  c  c} 
\ctoprule
\crowcolor\!\!Vector\!\!\!\!&${\cal B}_\mu$\!\!\!\!&${\cal B}_\mu^1$\!\!\!\!&${\cal W}_\mu$\!\!\!\!&${\cal W}_\mu^1$\!\!\!\!&${\cal G}_\mu$\!\!\!\!&${\cal G}_\mu^1$\!\!\!\!&${\cal H}_\mu$\!\!\!\!&${\cal L}_\mu$\!\!\!\!\\
\cmrule
\!\!Irrep\!\!\!\!&$\left(1,1\right)_0$\!\!\!\!&$\left(1,1\right)_1$\!\!\!\!&$\left(1,\Adj\right)_0$\!\!\!\!&$\left(1,\Adj\right)_1$\!\!\!\!&$\left(\Adj,1\right)_0$\!\!\!\!&$\left(\Adj,1\right)_1$\!\!\!\!&$\left(\Adj,\Adj\right)_{0}$\!\!\!\!&$\left(1,2\right)_{-\frac 32}$\!\!\!\!\\
\cbottomrule
&&&&&&&&\\
\ctoprule
\crowcolor\!\!Vector\!\!\!\!&${\cal U}_\mu^2$\!\!\!\!&${\cal U}_\mu^5$\!\!\!\!&${\cal Q}_\mu^1$\!\!\!\!&${\cal Q}_\mu^5$\!\!\!\!&${\cal X}_\mu$\!\!\!\!&${\cal Y}_\mu^1$\!\!\!\!&${\cal Y}_\mu^5$\!\!\!\!&\\
\cmrule
\!\!Irrep\!\!\!\!&$\left(3,1\right)_{\frac 23}$\!\!\!\!&$\left(3,1\right)_{\frac 53}$\!\!\!\!&$\left(3,2\right)_{\frac 16}$\!\!\!\!&$\left(3,2\right)_{-\frac 56}$\!\!\!\!&$\left(3,\Adj\right)_{\frac 23}$\!\!\!\!&$\left(\bar 6,2\right)_{\frac 16}$\!\!\!\!&$\left(\bar 6,2\right)_{-\frac 56}$\!\!\!\!&\\
\cbottomrule
\end{tabular}
}
\label{table:newvectors}
\end{center}
\end{table}

These representations contain Z$^\prime$ and W$^\prime$ vectors, gluon-like bosons, diquarks, leptoquarks, and other possibilities. All new vector bosons in arbitrary models are contained in this table, as long as they can be singly produced. Note that this excludes vectors that are odd under some R, T or KK parity. 
For phenomenological purposes, it is not important whether the new vector bosons are the gauge bosons of a broken extended gauge group or not. Nevertheless, it is interesting to note that all the types of vector bosons in Table~\ref{table:newvectors} can in principle be obtained as the gauge bosons of an extended gauge group broken down to the SM.  

To illustrate the power of the complete SM gauge invariance, as opposed to simple conservation of electric charge, let us study briefly two examples that are often included in electroweak fits and direct searches. First, consider charge $\pm 1$ vector bosons, and assume that they have sizable couplings to both leptons and quarks and moreover that there are no light right-handed neutrinos. It turns out that there is only one possible vector irreducible representation with these properties: $\mathcal{W}_\mu$. It couples only to the left-handed SM fermions, just as the SM $\SU{2}_L$ gauge boson. Thus, the charged components of this multiplet  form a sequential W$^\prime$. But these fields necessarily come together with the neutral component of the triplet, a Z$^\prime$ boson. This simple fact is usually not taken into account in collider searches, even if it is a model-independent consequence of the $\SU{2}_L$ gauge invariance of SM extensions. 

As a second example, consider the case of a sequential Z$^\prime$ boson, with couplings proportional to the ones of the SM $Z$ boson. This vector has different couplings to the two components of the $\SU{2}_L$ doublets, so it cannot be a singlet under the SM group. Nevertheless, it can arise after electroweak symmetry breaking as a mixture of a singlet vector and the third component of a vector in the adjoint of $\SU{2}_L$. This is the case of models with a replica of the SM gauge group, or in extra dimensions, but the mechanism is more general.  Gauge invariance implies that the sequential Z$^\prime$ boson necessarily comes together with a pair of charged vectors W$^\prime$ and another neutral vector $\gamma^\prime$, which couples just like the photon. All these new fields have similar masses. Clearly, these extra vectors have the very same structure of the SM gauge bosons.

Once the field content of the theory has been established, we can proceed to construct the most general renormalizable theory invariant under $\SU{3}_C\otimes \SU{2}_L \otimes \U{1}_Y$. The Lagrangian has the form
\be
\mathcal{L}=\mathcal{L}_\SM + \mathcal{L}_V + \mathcal{L}_{V-\SM} + ~\mbox{nonlinear,} \label{Lagrangian}
\ee
where $\mathcal{L}_\SM$ is the SM Lagrangian, $\mathcal{L}_V$ contains the quadratic terms for the heavy vector bosons (with kinetic terms covariantized with respect to the SM group) and $\mathcal{L}_{V-\SM}$ contains the possible interaction or kinetic terms formed as products of SM fields and a single vector field. Mass mixing terms of SM and new vectors are forbidden by gauge invariance.\footnote{There are, nevertheless, interactions with the Higgs doublet that give rise to mass mixing of the $Z$ and $W$ bosons with the new vectors when the electroweak symmetry is broken.} Finally, 
``nonlinear'' in Eq.~(\ref{Lagrangian}) refers to interaction terms that are nonlinear in the heavy vector fields. As we have argued above, those terms can be safely neglected. 
  
The quadratic terms for the new vector bosons are given by
\be
{\cal L}_V= - \sum_V \eta_V \left(\frac{1}{2} D_\mu V_{\nu}^\dagger D^\mu V^{\nu}-\frac{1}{2}D_\mu V_{\nu}^\dagger D^\nu V^{\mu}+\frac{1}{2}M_{V}^2 V_\mu^\dagger V^\mu \right),
\label{VLag}
\ee
The sum is over all new vectors $V$, which can be classified into the different irreducible representations of Table~\ref{table:newvectors}. We set $\eta_V=1 \;(2)$ when $V$ is in a real (complex) representation, in order to use the usual normalization. Note that we have written explicit mass terms for the new vectors. The masses can arise, in particular, from vacuum expectation values of extra scalar fields. In writing \refeq{VLag}, we have chosen a basis with diagonal, canonically normalized kinetic terms and diagonal masses. 
The couplings of the new vectors to the SM are described by
\be
\mathcal{L}_{V-\SM}= - \sum_V \frac{\eta_V}{2}\left({V^\mu}^\dagger J^V_\mu +\hc\right).
\ee
The vector currents $J^V_\mu$ above have the form
\be
J^V_\mu = \sum_k g_V^k j^{Vk}_\mu,
\ee
where $g_V^k$is a coupling constant and $j^{Vk}_\mu$ is a vector operator of scaling dimension 3 in the same representation as $V$. Actually, the different currents that can be built with the SM fields determine the possible representations of the extra vectors. We can distinguish three kinds of SM currents:
\begin{itemize}
\item {\em With two fermions\/}. Schematically, $j^{V\psi_1\psi_2}_\mu = [\overline{\psi_1} \otimes \gamma_\mu \psi_2]_{R_V}$, with $\psi_1$, $\psi_2$ (different in principle) fermion multiplets, $R_V$ the representation of $V$ and $\otimes$ a product of representations.
\item {\em With two scalars and a covariant derivative\/}: $j^{V\phi}_\mu = [\Phi^\dagger \otimes D_\mu \phi]_{R_V}$, where $\Phi$ denotes either the scalar doublet $\phi$ or its form $\tilde{\phi}$.
\item {\em With a gauge boson and two covariant derivatives\/}: $j^{\mathcal{A}}_\mu = D^\nu A_{\nu\mu}$.
\end{itemize}
The couplings to currents of the third type induce a kinetic mixing of the SM gauge bosons $A$ with the heavy vectors $\mathcal{A}$. It turns out that the corresponding terms in $\mathcal{L}_\mathrm{V-SM}$ are redundant and can be eliminated by field redefinitions.

The currents $J^V$ for all possible vectors $V$ and the most general couplings have been given in Ref.\ \cite{Jorge}. Let us write, as an example, the current for the vector boson $\mathcal{B}^1_\mu$:
\be
J^{{\cal B}^1}_\mu=\left(g^{du}_{{\cal B}^1}\right)_{ij}\overline{d_R^i}\gamma_\mu u^j_R+g_{{\cal B}^1}^\phi i D_\mu \phi^Ti\sigma_2\phi.
\ee
We see that in this case, up to flavour indices, there are just two independent couplings: the first one induces right-handed charged currents, while the second one gives rise to a mixing with the SM W$^{\pm}$ boson, which modifies the $\rho$ parameter. 

\begin{figure}
\begin{center}
\includegraphics[width=0.6\textwidth]{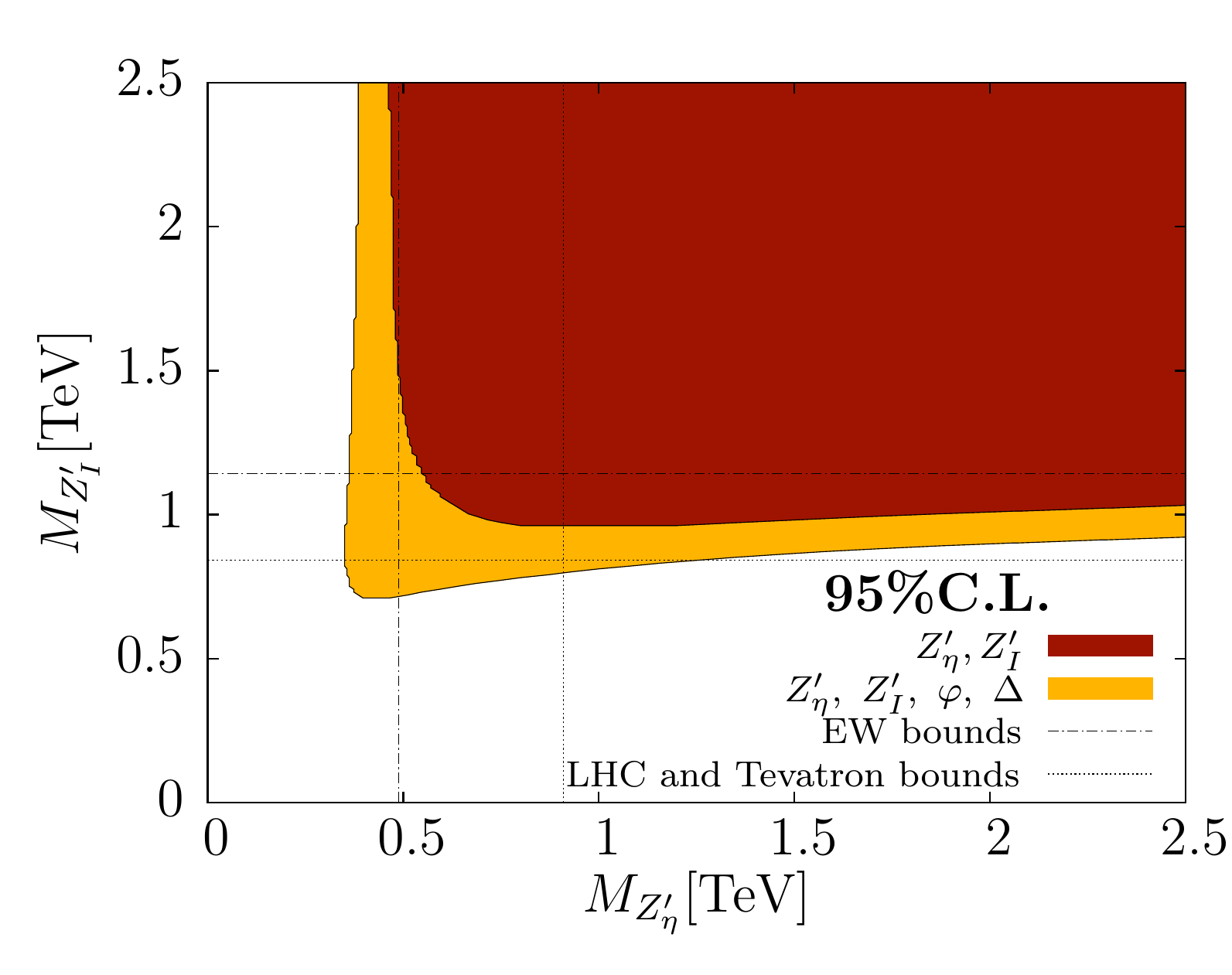}
\caption{95\% C.L.\ regions in the $M_{Z_I^\prime}-M_{Z_\eta^\prime}$ plane for a two-Z$^\prime$ global fit with and without extra scalars $\phi$ (singlet) and $\Delta$ (isotriplet).
\label{fig:cancellations}}
\end{center}
\end{figure}

The model-independent limits from EWPD on all the extra vectors have been presented in Ref.\cite{Jorge}. On the other hand, Tevatron and LHC are placing better and better bounds on many of the vector bosons, especially those that have sizable couplings to both quarks and leptons. Here we shall just point out that the indirect limits can be relaxed in some cases with additional new particles, as has been systematically discussed in Ref.\ \cite{cancellations}. As an example of this, let us show the constraints on models with two kinds of Z$^\prime$ bosons that appear in $E_6$ GUT, a Z$^\prime_I$ and a Z$^\prime_\eta$ (see \cite{Langacker:2008yv}), plus optional extra scalar singlets $\phi$ and triplets $\Delta$. In Fig. \ref{fig:cancellations} we plot the 95\% C.L.\ regions, together with the electroweak and direct limits for the cases of just one Z$^\prime$. 
We see that the interplay of the different particles weakens the limits for individual Z$^\prime$, and enlarges the parameter space available for discovery at the LHC.


\section{A few consequences of extra vector bosons}

We start by discussing the implications of new vector bosons on the value of the mass of the Higgs boson. Some of the new vectors modify the $\rho$ parameter at tree level. In particular, the $\mathcal{B}_\mu$ and $\mathcal{W}^1_\mu$ representations can produce a shift that counteracts the effect of loops with a heavy Higgs (relative to the ones with a light Higgs). 
\begin{figure}
\begin{center}
\includegraphics[width=0.6\textwidth]{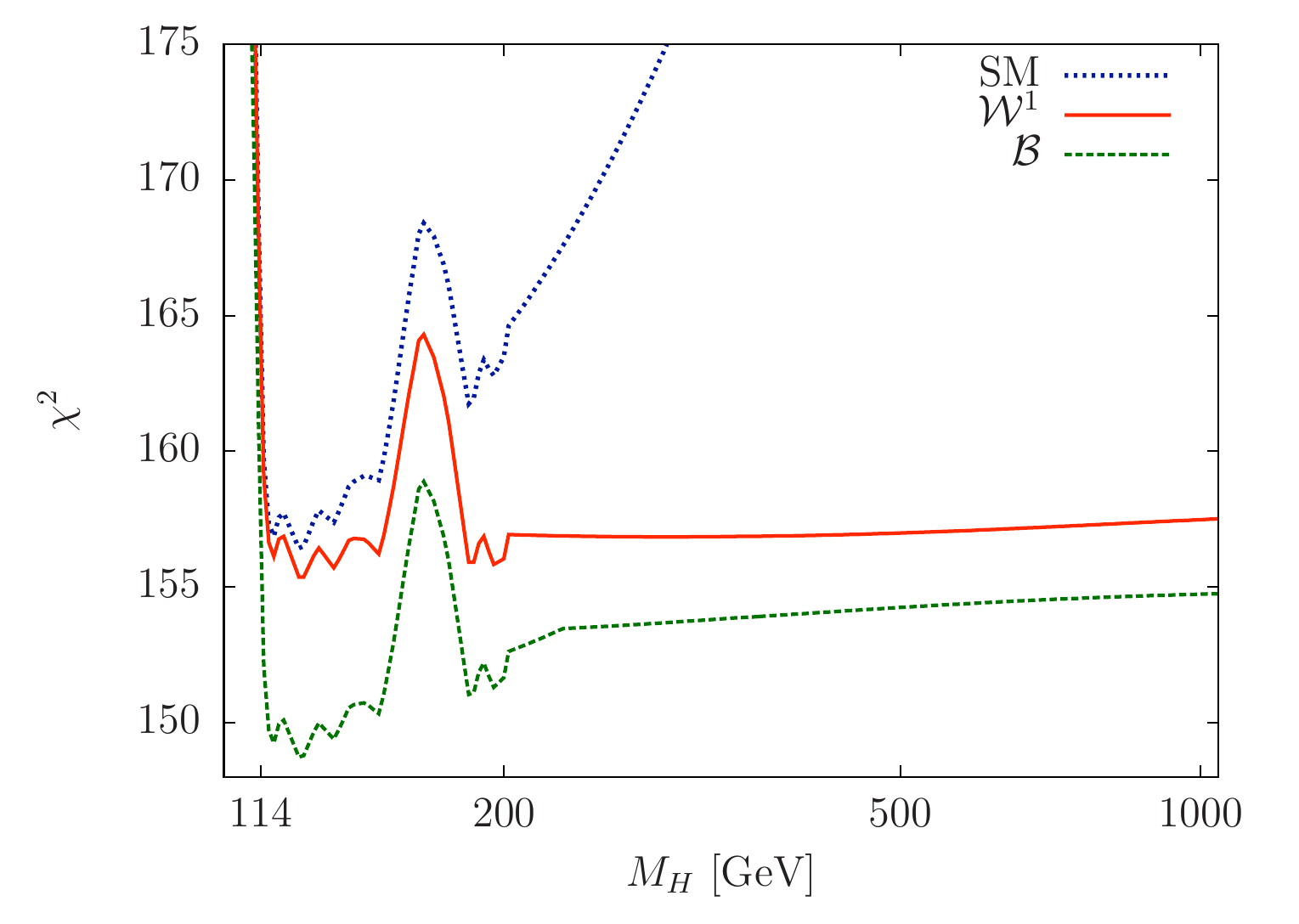}
\caption{$\chi^2$ of best fit as a function of the Higgs mass $M_H$ for the SM and extensions with $\mathcal{B}_\mu$ and $\mathcal{W}^1_\mu$ vectors.
\label{fig:higgsmass}}
\end{center}
\end{figure}
In Fig.~\ref{fig:higgsmass} we plot the value of the $\chi^2$ of a global fit in three scenarios: the SM, an extension with a singlet  $\mathcal{B}_\mu$ and an extension with a hypercharged triplet  $\mathcal{W}^1_\mu$. The fit includes EWPD, LEP 2 data and information from direct searches of the Higgs boson at Tevatron.  The value of $\chi^2$ over the number of degrees of freedom is comparable in all three scenarios for a light Higgs, $M_H \sim 125 \, \mathrm{GeV}$. However, it is clear from the figure that for a heavy Higgs both extensions with extra vector bosons are clearly favored over the SM hypothesis. In other words, if a heavy Higgs boson were found at the LHC, it would be an indication of physics beyond the SM, possibly in the form of new vector bosons $\mathcal{B}_\mu$ or $\mathcal{W}^1_\mu$.

\begin{figure}
\begin{center}
\includegraphics[width=0.6\textwidth]{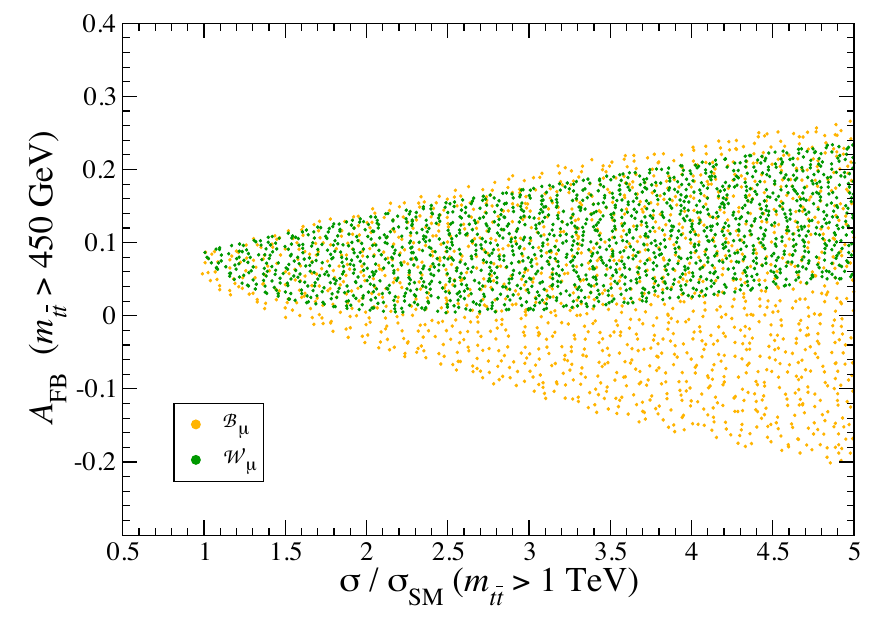}
\caption{Allowed regions for the Tevatron $t\bar{t}$ asymmetry and the $t\bar{t}$ tail at LHC for heavy  $\mathcal{B}_\mu$ and $\mathcal{W}_\mu$ bosons.
\label{fig:asymmetry}}
\end{center}
\end{figure}
Let us now move to the possible impact of extra vector bosons in top physics. Tevatron and the LHC are studying in great detail the pair production of top quarks. Interestingly, the CDF collaboration has measured a value of the $t\bar{t}$ forward backward asymmetry $A_{FB}$ (at high invariant mass) that is more than three sigmas away from the SM prediction~\cite{CDF}. Having a look at Table \ref{table:newvectors} and at the couplings in Ref.\ \cite{Jorge}, it is easy to see that the vector bosons contributing to top pair production are:
\begin{itemize}
\item $\mathcal{B}_\mu$, $\mathcal{W}_\mu$, $\mathcal{G}_\mu$ and $\mathcal{H}_\mu$, in the s and/or t channels;
\item $\mathcal{B}^1_\mu$, $\mathcal{G}^1_\mu$, in the t channel;
\item $\mathcal{Q}^1_\mu$, $\mathcal{Q}^5_\mu$, $\mathcal{Y}^1_\mu$ and $\mathcal{Y}^5_\mu$, in the u channel.
\end{itemize}
All of these vector bosons, except $\mathcal{Y}^1_\mu$ and $\mathcal{Y}^1_\mu$, could give rise to the observed excess. However, this is not that simple, as there are important constraints as well. The more robust ones arise from $t\bar{t}$ production itself: the total cross section and its distribution as a function of the invariant mass. In Ref. \cite{ASPV}, Juan Antonio Aguilar-Saavedra and the author have studied the effect of these vector bosons (and also of general scalars) over the tail in $t\bar{t}$ production at high invariant masses at the LHC. The result is that, with the exception of very light particles or particular couplings in the case of a gluon-like vector $\mathcal{G}$, these explanations of the excess in $A_{FB}$ can be ruled out by LHC data with the luminosity already collected. In Fig.\ \ref{fig:asymmetry} we show, for example, the relation between the predictions for $A_{FB}$ and the LHC tail for the case of heavy $\mathcal{B}_\mu$ and $\mathcal{W}_\mu$ vector bosons. The different points scan all the allowed values of the couplings of these vector bosons to the $t$, $u$ and $d$ quarks.

\section{Conclusions}
Many beautiful models incorporating new physics have been constructed in the last thirty years, guided by different theoretical problems of the SM. We do not know which, if any, of these theories is realized in our Universe. Therefore, now that the LHC is running---and performing extremely well!---it seems wise not to trust particular models, but carry out model independent analyses of new physics. These studies can guide the experimental searches and give shape to the constraints. The most general model-independent formalism, effective Lagrangians, is not valid when the new particles can be produced. We advocate instead a scan over all possible particles that can give observable effects, allowing for completely general couplings. This is not an impossible mission once the possibilities are strongly restricted by the principles of gauge invariance and power counting. The result is a natural and convenient parameterization from both the theoretical and experimental points of view. In this talk, we have illustrated this program for the case of extra spin 1 particles. 

\section*{Acknowledgments}
It is a pleasure to thank Paco del Aguila, Juan Antonio Aguilar-Saavedra, Jorge de Blas and Paul Langacker for a fruitful collaboration, and the organizers of the Rencontres de Moriond for a very enjoyable meeting. This work has been partially supported by projects FPA 2010-17915 (MICINN), FQM 101 and FQM 437 (Junta de Andaluc\'{\i}a).

\end{document}